\documentclass[12pt]{article}
\usepackage{epsfig}

\newcommand{\mysection}{\setcounter{equation}{0}\section}

\def\beq{\begin{equation}}
\def\eeq{\end{equation}}
\def\beqa{\begin{eqnarray}}
\def\eeqa{\end{eqnarray}}

\newlength{\dinwidth} \newlength{\dinmargin}
\begin{document}

\begin{center}
{\Large \bf Total and differential cross sections for Higgs and top-quark production}
\end{center}
\vspace{2mm}
\begin{center}
{\large Nikolaos Kidonakis{\footnote{Presented at ICNFP2018, Kolymbari, Crete, Greece, 4-12 July 2018}}}\\
\vspace{2mm}
{\it Department of Physics, Kennesaw State University,\\
Kennesaw, GA 30144, USA}
\end{center}

\begin{abstract}
I present theoretical calculations for Higgs-boson and top-quark production, including high-order soft-gluon corrections. I discuss charged-Higgs production in association with a top quark or a $W$ boson, as well as single-top and top-antitop production. Total cross sections as well as transverse-momentum and rapidity distributions of the top quark or the Higgs boson are presented for various LHC energies.
\end{abstract}

\mysection{Introduction}

The study of the top quark and of Higgs bosons in the Standard Model and beyond are two major active areas of fundamental physics and its exporation at the LHC. To make the most of the physics program at the LHC we need to be able to predict theoretically the cross sections for processes involving top quarks and Higgs bosons, including processes beyond the Standard Model.  Perturbative QCD corrections are typically large for these processes and they are needed to reduce dependence of the cross sections on factorization and renormalization scales. 

Many processes have now been calculated to NLO, and some to NNLO. The complexity of the calculations increases enormously with each order. It is therefore important to identify the sources of specific contributions to the cross sections, and whether they are numerically dominant and can be calculated with alternative techniques. It turns out that soft-gluon corrections, i.e. radiative corrections calculated in the eikonal approximation where the gluons are low-energy, dominate the cross sections for many processes at LHC and Tevatron energies. It is thus important and meaningful to calculate these corrections. See Ref. \cite{NKtop} for a review.

These soft-gluon corrections can be formally resummed in moment space to all orders in the perturbative series by using factorization of the cross section into different functions that describe the behavior of hard, collinear, and soft quanta, and renormalization-group evolution of these functions. However, physical predictions for resummed cross sections need some kind of method or prescription to avoid Landau singularities in the resummed result, and the record of such prescriptions has been rather poor in that they typically grossly underestimate the numerical contribution of the corrections. However, fixed-order expansions of the resummed cross sections bypass such concerns. 

Expansions of resummed cross sections have been given at NLO, NNLO, and even N$^3$LO for numerous top-quark and Higgs processes \cite{NKtop}. The soft-gluon corrections at NLO typically are excellent approximations to the complete NLO corrections. In cases where the complete NNLO corrections are known, it is also found that the soft-gluon corrections at NNLO are also very good approximations. Even higher-order soft-gluon corrections can provide additional significant contributions.

In this presentation, I discuss the latest results with soft-gluon corrections for various processes involving charged Higgs bosons and top quarks. In particular I discuss $tH^-$ production, $H^-W^+$ production, single-top quark production in the $t$- and $s$-channels and $tW$ production, $tZ$ production via anomalous top-quark couplings, and $t{\bar t}$ production. Soft-gluon corrections are very important in all these cases and they approximate exact results very well.
We resum these soft corrections at next-to-next-to-leading logarithm (NNLL) accuracy for the double-differential cross section, and we use the resummed cross section as a generator of finite-order expansions to provide  approximate NNLO (aNNLO) and approximate N$^3$LO (aN$^3$LO) predictions for cross sections and differential distributions.

\mysection{Higher-order soft-gluon corrections}

We consider partonic processes for charged-Higgs production
$$
f_{1}\, + \, f_{2}\, \rightarrow \, H^-\, + \, X 
$$
and for top-quark production
$$
f_{1}\, + \, f_{2}\, \rightarrow \, t\, + \, X \, .
$$
In addition to the usual kinematical variables $s$, $t$, $u$,  
we define a threshold variable $s_4=s+t+u-\sum m^2$. At partonic threshold $s_4 \rightarrow 0$,
and the soft-gluon corrections  are of the form $[\ln^k(s_4/m_H^2)/s_4]_+$
for charged-Higgs production, and $[\ln^k(s_4/m_t^2)/s_4]_+$ for top-quark production,
with $k \le 2n-1$ at order $\alpha_s^n$.
We resum these soft corrections for the double-differential cross section in $t$ and $u$, 
or equivalently in $p_T$ and rapidity.

To derive soft-gluon resummation, we take moments of the partonic cross section with moment variable $N$,  
${\hat \sigma}(N)=\int (ds_4/s) \; e^{-N s_4/s} {\hat \sigma}(s_4)$, 
and write a factorized expression for the cross section in $4-\epsilon$ dimensions
$$
\sigma(N,\epsilon)
= H_{IL}\left(\alpha_s(\mu_R)\right) \, 
S_{LI}\left(\frac{m}{N \mu_F},\alpha_s(\mu_R) \right)
\prod  J_{\rm in} \left(N,\mu_F,\epsilon \right)
\prod J_{\rm out} \left(N,\mu_F,\epsilon \right)
$$ 
where $m$ denotes the charged-Higgs or top-quark masses, depending on the process, and $\mu_F$ and $\mu_R$ are the factorization and renormalization scales. 
$H_{IL}$ is the hard function and $S_{LI}$ is the soft function, both of them matrices in general in the space of color exchanges, and the $J_{\rm in}$ and $J_{\rm out}$ collect collinear and soft-gluon corrections from the incoming and outgoing partons.
$S_{LI}$ satisfies the renormalization group equation
$$
\left(\mu \frac{\partial}{\partial \mu}
+\beta(g_s)\frac{\partial}{\partial g_s}\right)\,S_{LI}
=-(\Gamma^\dagger_S)_{LK}S_{KI}-S_{LK}(\Gamma_S)_{KI}
$$
where $\Gamma_S$ is the soft anomalous dimension,  
thus resulting in the exponentiation of logarithms of $N$. 
At NNLL accuracy we need two-loop soft anomalous dimensions.

Resummed expressions follow from the evolution of the soft function, as determined by the above renormalization group equation, as well as of the $J$ functions in the factorized expression.

\mysection{Charged Higgs production}

We begin with the associated production of a top quark and a charged Higgs boson in the MSSM or other two-Higgs-doublet models \cite{NKtH}.
The lowest-order cross section for the process $bg \rightarrow t H^-$ is proportional to
$\alpha \alpha_s (m_b^2\tan^2 \beta+m_t^2 \cot^2 \beta)$
where $\tan \beta=v_2/v_1$ is the ratio of the vacuum expectation values of the two Higgs doublets.
The soft anomalous dimension for the process is 
$\Gamma_S^{bg \rightarrow tH^-}=(\alpha_s/\pi) \Gamma_S^{(1)}+(\alpha_s/\pi)^2 \Gamma_S^{(2)}+\cdots$, with
$$
\Gamma_S^{(1)}=C_F \left[\ln\left(\frac{m_t^2-t}{m_t\sqrt{s}}\right)
-\frac{1}{2}\right] +\frac{C_A}{2} \ln\left(\frac{m_t^2-u}{m_t^2-t}\right)
$$
and 
$$
\Gamma_S^{(2)}=\left[C_A \left(\frac{67}{36}-\frac{\zeta_2}{2}\right)-\frac{5}{18} n_f\right] \Gamma_S^{(1)} +C_F C_A \frac{(1-\zeta_3)}{4} \, .
$$

The analytical expressions for the soft-gluon corrections through NNLO for $tH^-$ production have been given in \cite{NKtH}. The approximate N$^3$LO (aN$^3$LO) soft-gluon corrections are:
\beqa
\frac{d^2{\hat \sigma}_{\rm aN^3LO}^{(3) \, bg \rightarrow tH^-}}{dt \, du}
&=&F_{\rm LO}^{bg \rightarrow tH^-} \frac{\alpha_s^3}{\pi^3} 
\left\{(C_F+C_A)^3 \left[\frac{\ln^5(s_4/m_H^2)}{s_4}\right]_+ \right.
\nonumber \\ &&
{}+\frac{5}{2}(C_F+C_A)^2\left[2C_F\ln\left(\frac{m_t^2-t}{m_t \sqrt{s}}\right)
-2C_F\ln\left(\frac{m_H^2-u}{m_H^2}\right)-C_F \right.
\nonumber \\ && 
{}+C_A\ln\left(\frac{m_t^2-u}{m_t^2-t}\right)
-2C_A \ln\left(\frac{m_H^2-t}{m_H^2}\right)
-(C_F+C_A) \ln\left(\frac{\mu_F^2}{s}\right)
\nonumber \\ && \left. \left.
{}-\frac{11}{9}C_A+\frac{2}{9}n_f\right] 
\left[\frac{\ln^4(s_4/m_H^2)}{s_4}\right]_+ 
+{\cal O}\left(\left[\frac{\ln^3(s_4/m_H^2)}{s_4}\right]_+ \right)\right\} 
\nonumber
\eeqa
where, for brevity, we do not show explicitly the lower powers of the logarithms.

\begin{figure}[h]
\centering
\includegraphics[width=80mm]{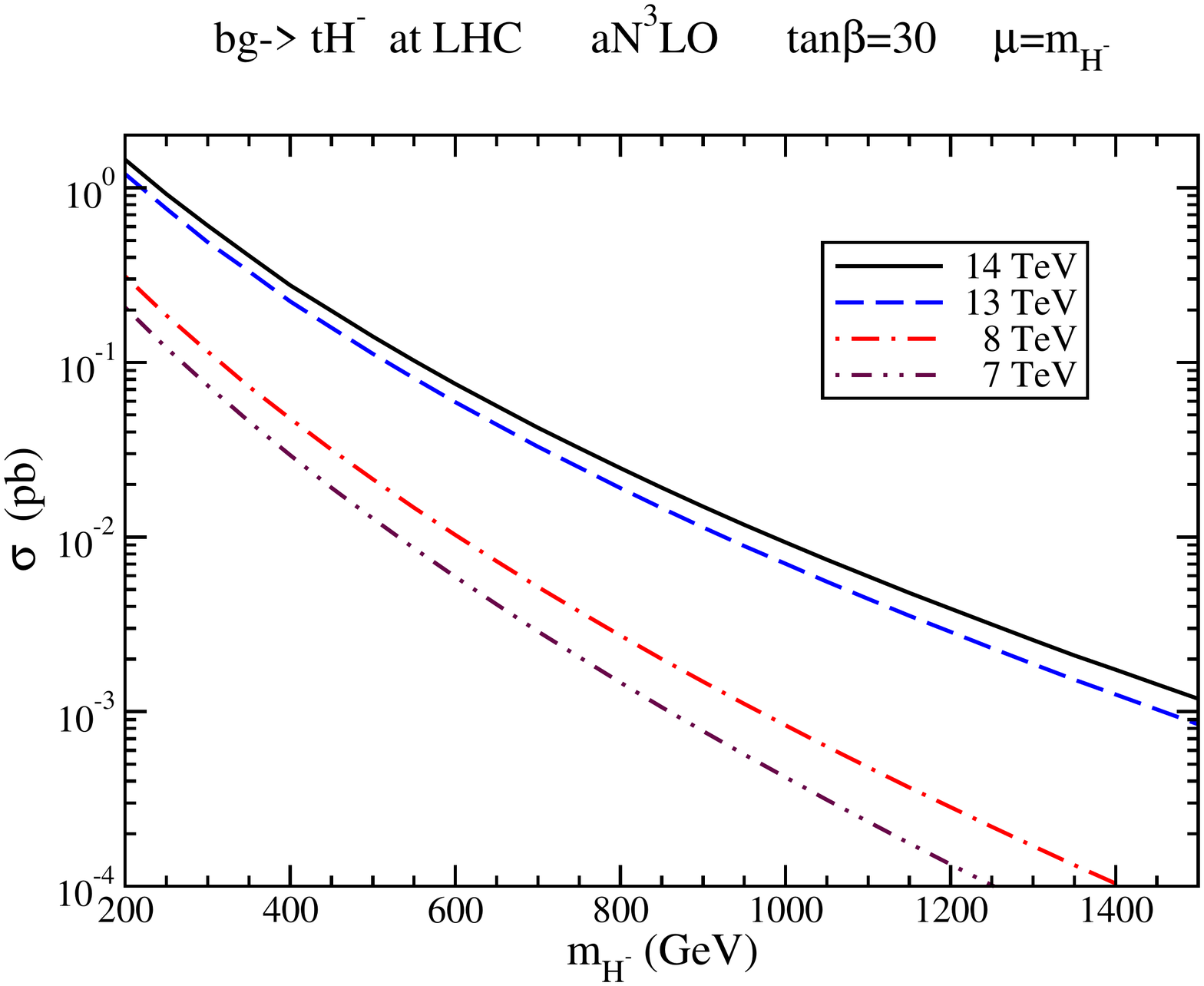}
\includegraphics[width=80mm]{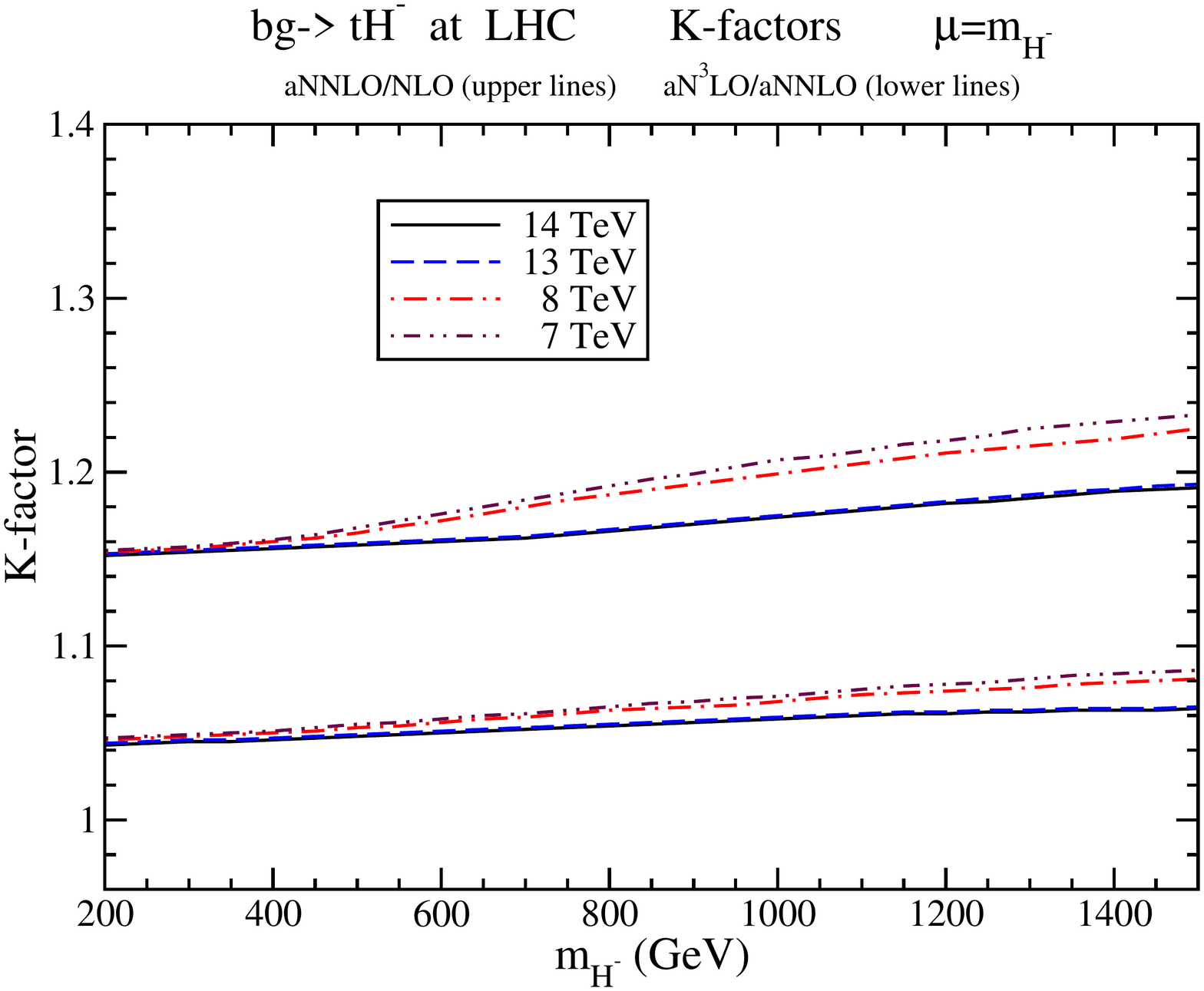}
\caption{(Left) aN$^3$LO total cross sections for $tH^-$ production. (Right) $K$-factors for $tH^-$ production.}
\label{tH}
\end{figure}

We now present the aN$^3$LO total cross sections at LHC energies. We use MMHT2014 NNLO pdf \cite{MMHT2014} for our numerical results. In the left plot of Fig. \ref{tH} we show the aN$^3$LO total cross section for $tH^-$ production, with $\tan\beta=30$, as a function of charged-Higgs mass at LHC energies of 7, 8, 13, and 14 TeV. The soft-gluon corrections are large for this process, as shown on the plot on the right. Top-quark $p_T$ and rapidity distributions in this process have also been presented in \cite{NKtH}.

\begin{figure}[h]
\centering
\includegraphics[width=80mm]{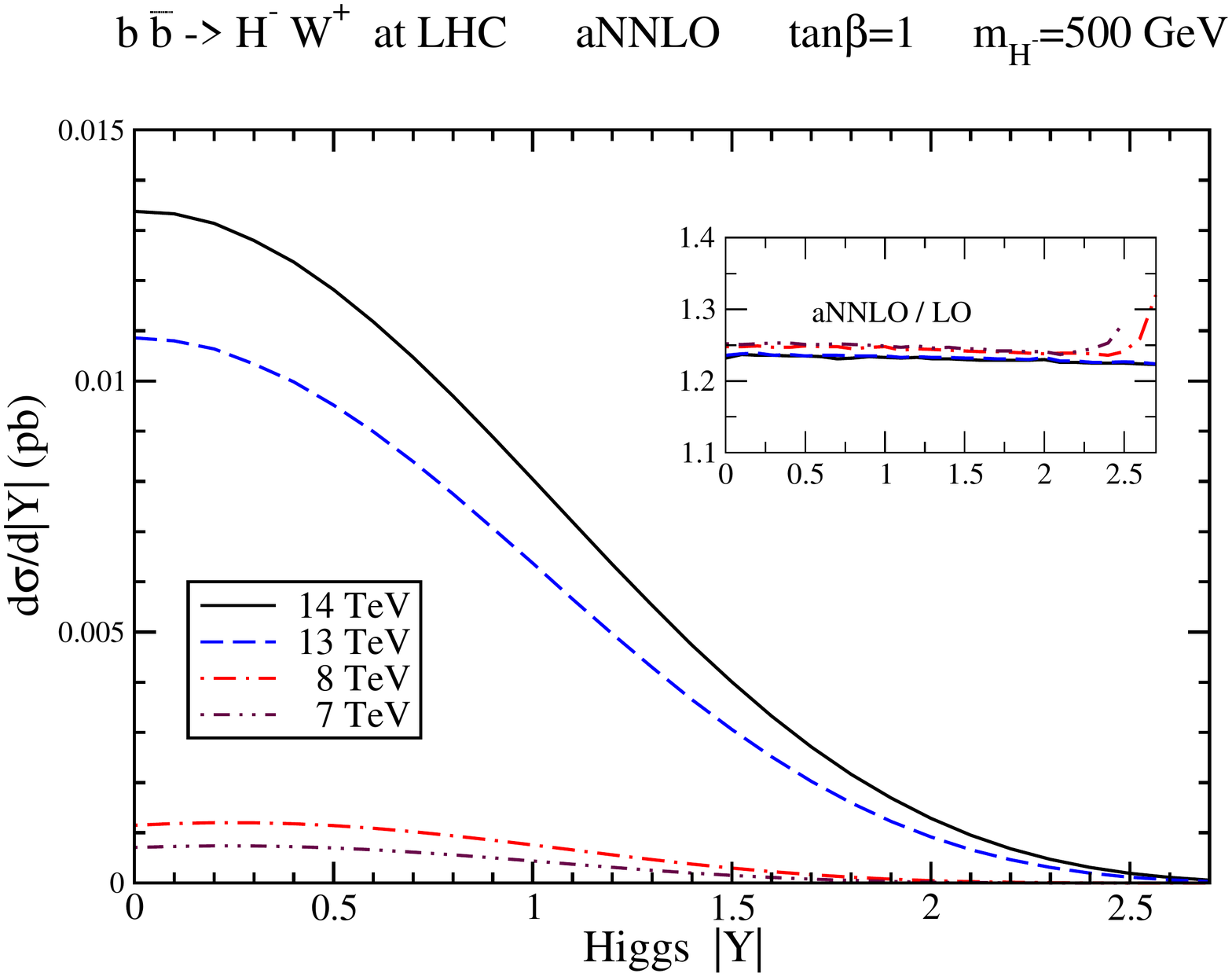}
\includegraphics[width=80mm]{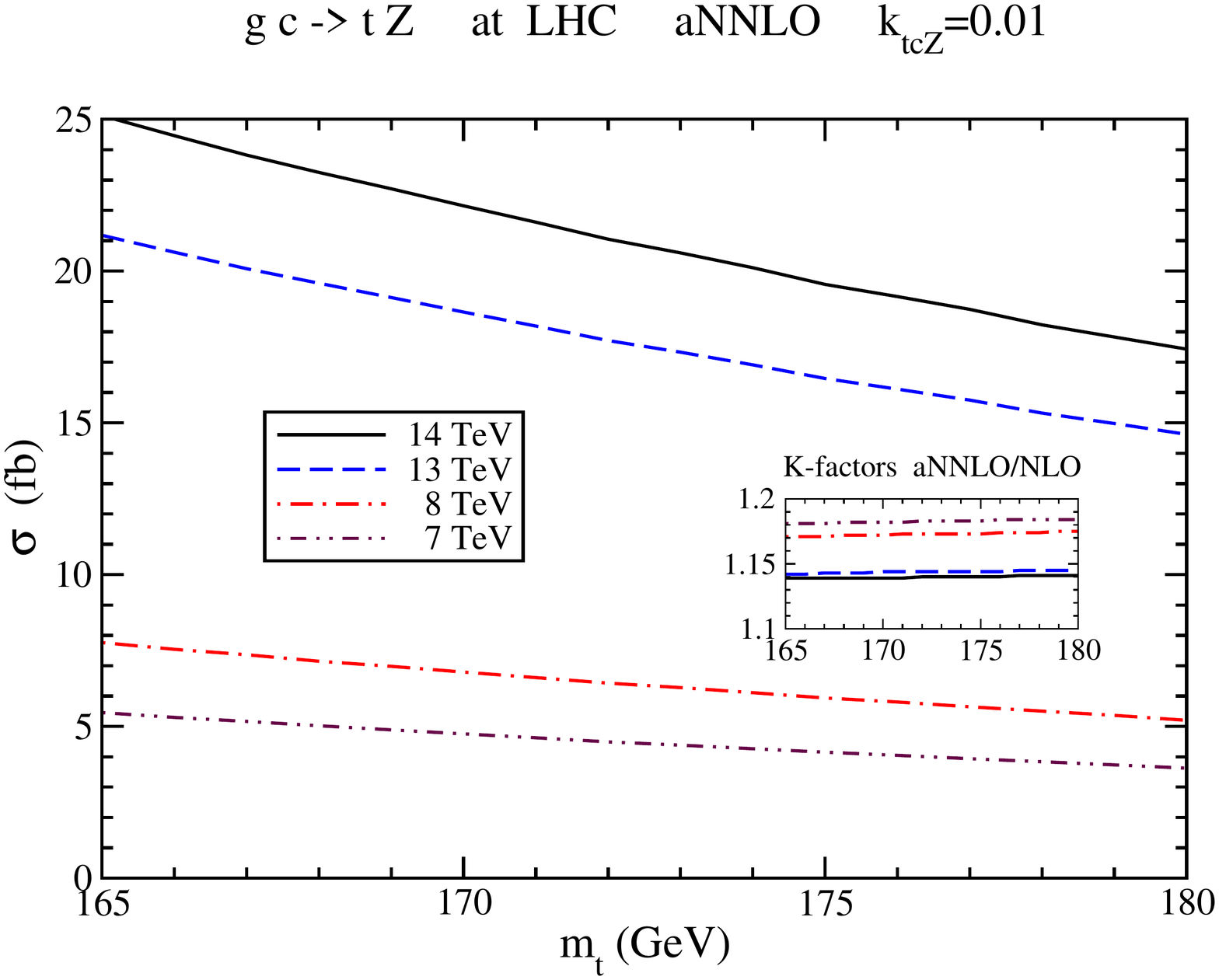}
\caption{(Left) The aNNLO rapidity distribution of the charged Higgs boson in $H^-W^+$ production at LHC energies. (Right) The aNNLO total cross section for the process $gc \rightarrow tZ$.}
\label{HWtZ}
\end{figure}

We next discuss $H^-W^+$ production via $b{\bar b} \rightarrow H^- W^+$ \cite{NKHW}. In the left plot of Fig. \ref{HWtZ} we show the aNNLO charged-Higgs rapidity distribution $d\sigma/d|Y|$, with $\tan\beta=1$ and $m_H=500$ GeV, at 7, 8, 13, and 14 TeV LHC energies with MMHT2014 NNLO pdf. The inset plot shows the aNNLO/LO $K$-factors. The soft-gluon corrections are clearly very significant. 

\mysection{$tZ$ production via anomalous couplings}

In physics beyond the Standard Model, top-quark production may occur via anomalous couplings of the top quark. Here, we discuss soft-gluon corrections in $tZ$ production with anomalous $t$-$q$-$Z$ couplings via the processes $gu \rightarrow tZ$ and $gc \rightarrow tZ$ \cite{NKtZ}. The complete NLO corrections \cite{NLOtZ} are very well approximated by the soft-gluon corrections at that order.

In the right plot of Fig. \ref{HWtZ} we show the aNNLO total cross section for $gc \rightarrow tZ$ as a function of top-quark mass at LHC energies of 7, 8, 13, and 14 TeV with CT14 pdf \cite{CT14}. The inset plot displays the $K$-factors, which show that the aNNLO corrections are large. The fact that these corrections significantly enhance the NLO cross section is an important theoretical input to setting experimental limits on the couplings \cite{CMStqZ,ATLAStqZ}. Similar results are found for the process $gu \rightarrow tZ$, and top-quark differential distributions for both processes have been presented in Ref. \cite{NKtZ}.

Related calculations have more recently been done for $t \gamma$ production via anomalous couplings in Ref. \cite{MFNK}, with similar findings on the importance of soft-gluon corrections.

\mysection{Single-top production}

Next, we discuss single-top production in the $t$-channel, $s$-channel, and via $tW$ production. These processes are now known at NNLO for the $t$-channel \cite{NNLOtch,BGYZ,BGZ} and the $s$-channel \cite{NNLOsch}, and at NLO for $tW$ production \cite{Zhu}. Soft-gluon resummation at NNLL has been performed for all channels \cite{NKsingletop,NKtW16}. For the $t$ and $s$ channels the soft anomalous dimension is a $2\times 2$ matrix in color space, while for $tW$ production it is the same as the one we presented for $tH^-$ production. We now present results with soft-gluon corrections at aNNLO \cite{NKsingletop} for the $t$ and $s$ channels, and at aN$^3$LO \cite{NKtW16} for $tW$ production.

\begin{figure}[h]
\centering
\includegraphics[width=85mm]{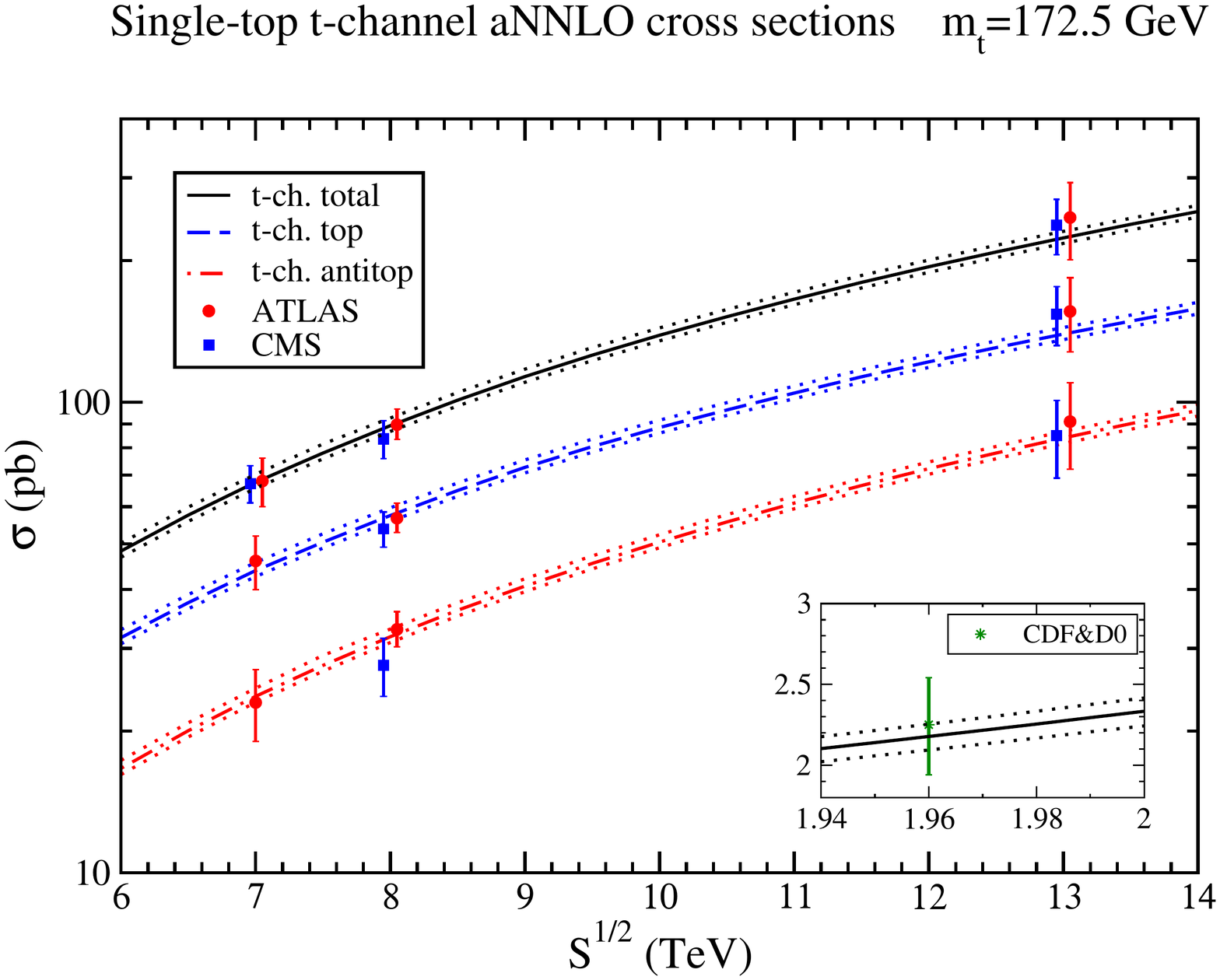}
\hspace{-3mm}
\includegraphics[width=85mm]{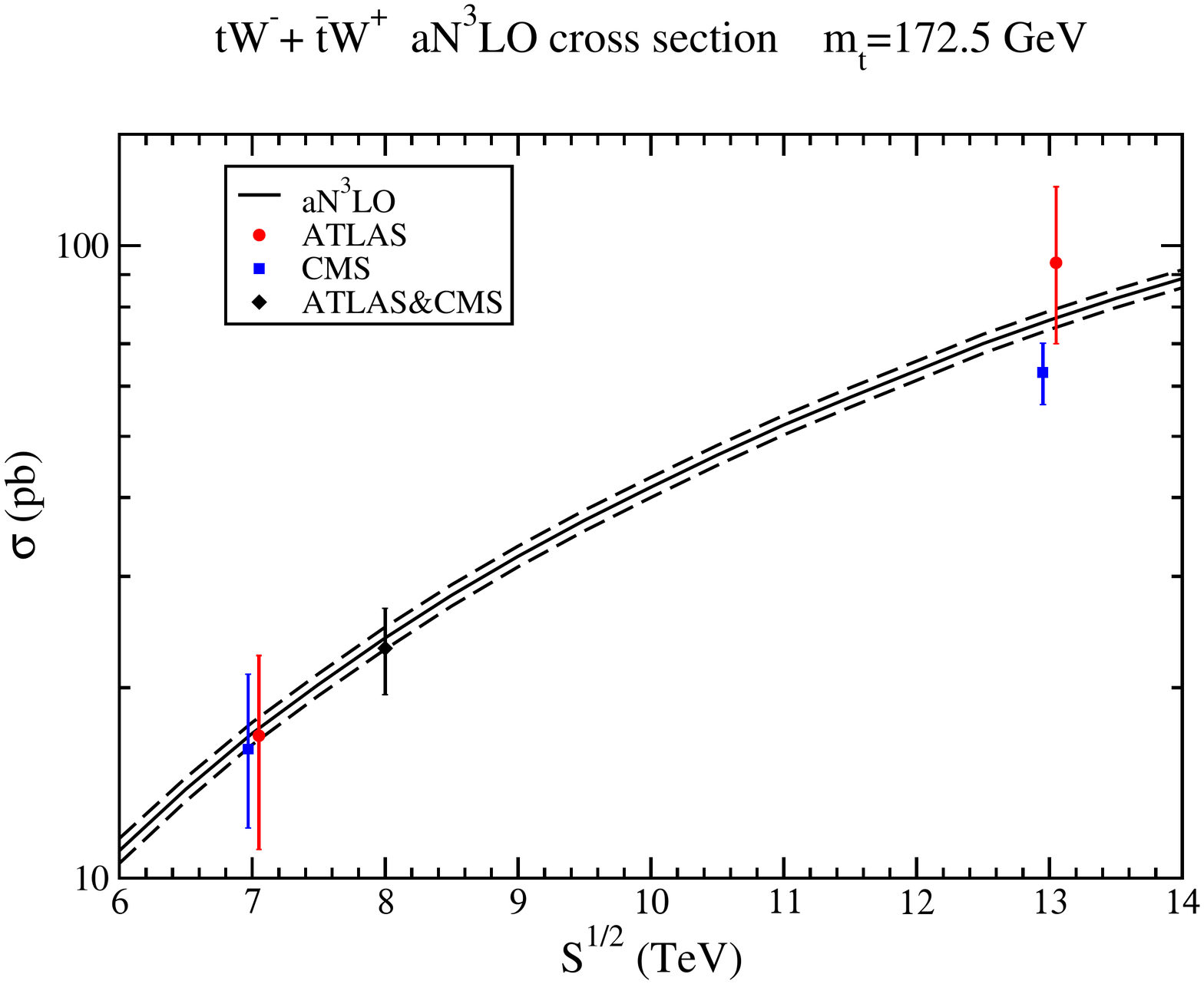}
\caption{(Left) Single-top $t$-channel aNNLO cross sections compared with CMS and ATLAS data at 7 TeV \cite{CMStch7,ATLAStch7}, 8 TeV \cite{CMStch8,ATLAStch8}, and 13 TeV \cite{ATLAStch13,CMStch13}, and with CDF and D0 combined data at 1.96 TeV \cite{CDFD0tch}. (Right) aN$^3$LO cross sections for $tW$ production compared to ATLAS and CMS data at 7 TeV \cite{ATLAStW7,CMStW7}, 8 TeV \cite{ATLASCMStW8}, and 13 TeV \cite{ATLAStW13,CMStW13}.}
\label{tchtW}
\end{figure}

In the left plot of Fig. \ref{tchtW} we show aNNLO results for $t$-channel cross sections, using MMHT2014 NNLO pdf \cite{MMHT2014}, at LHC and (inset) at Tevatron energies. Results are shown separately for the single-top cross section, the single-antitop cross section, and their sum. We find excellent agreement of the aNNLO predictions with all data from the LHC and the Tevatron.
The aNNLO normalized top-quark $p_T$ distributions also describe the available data quite well \cite{QCDwork18}.

We next discuss $tW$ production at aN$^3$LO. In the right plot of Fig. \ref{tchtW} we show the total $tW^-$+${\bar t}W^+$ cross section as a function of energy. We observe very good agreement with LHC data at 7, 8, and 13 TeV energies.

Finally, the theoretical predictions for $s$-channel production at aNNLO are in good agreement with available LHC and Tevatron data as shown in \cite{QCDwork18}.

\mysection{Top-antitop pair production}

Top-antitop pair poduction is the dominant mode at both the Tevatron and the LHC. The theoretical state-of-the-art is currently aN$^3$LO \cite{NKtt}. The soft anomalous dimensions are known at two loops and they are $2 \times 2$ matrices in color space for the $q{\bar q}\rightarrow t{\bar t}$ channel, and $3 \times 3$ matrices for the $gg\rightarrow t{\bar t}$ channel.

\begin{figure}[h]
\centering
\includegraphics[width=10cm]{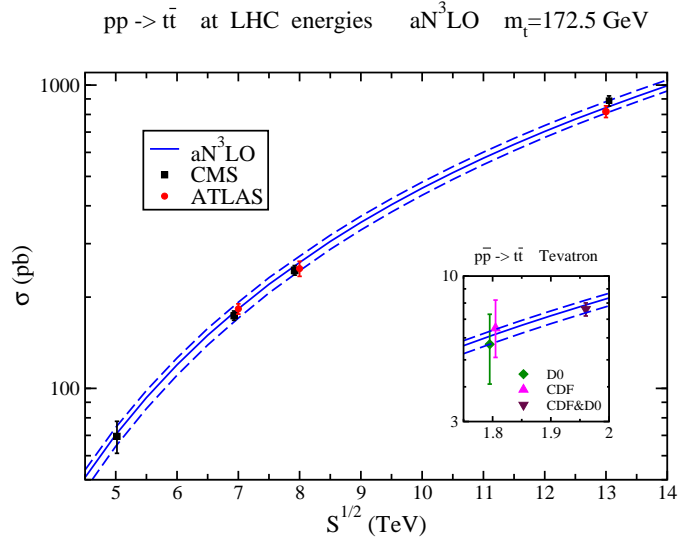}
\caption{Top-antitop aN$^3$LO cross sections compared with CMS data at 5.02 TeV \cite{CMS5.02} and with ATLAS and CMS data at 7 TeV \cite{ATLAStt7,CMStt7and8}, 8 TeV \cite{CMStt7and8,ATLAStt8}, and 13 TeV \cite{ATLAStt13,CMStt13} LHC energies. The inset shows the aN$^3$LO cross section compared with CDF \cite{CDF1.8} and D0 \cite{D01.8} data at 1.8 TeV, and CDF\&D0 combined data \cite{CDFD01.96} at 1.96 TeV Tevatron energy.}
\label{tt}
\end{figure}

The soft-gluon corrections have been known at NLO and NNLO for total and differential cross sections for some time. These corrections are large and dominant, and they provide excellent approximations to the complete QCD corrections at both NLO and NNLO \cite{NKtop}. The further corrections at aN$^3$LO are still significant, and they need to included in theoretical predictions for improved accuracy and smaller theoretical uncertainty.

Figure \ref{tt} displays the aN$^3$LO top-antitop cross sections at LHC and Tevatron energies using MMHT2014 NNLO pdf. There is data from the Tevatron at 1.8 TeV and 1.96 TeV energies, and from the LHC at 5.02, 7, 8, and 13 TeV energies. The theoretical curves describe the data at all energies remarkably well. The fact that aN$^3$LO theory has predicted and agrees with the data over such a wide energy range is highly significant. The soft-gluon corrections are important both in enhancing the cross section and in reducing its dependence on renormalization and factorization scales.

\begin{figure}[h]
\centering
\includegraphics[width=85mm]{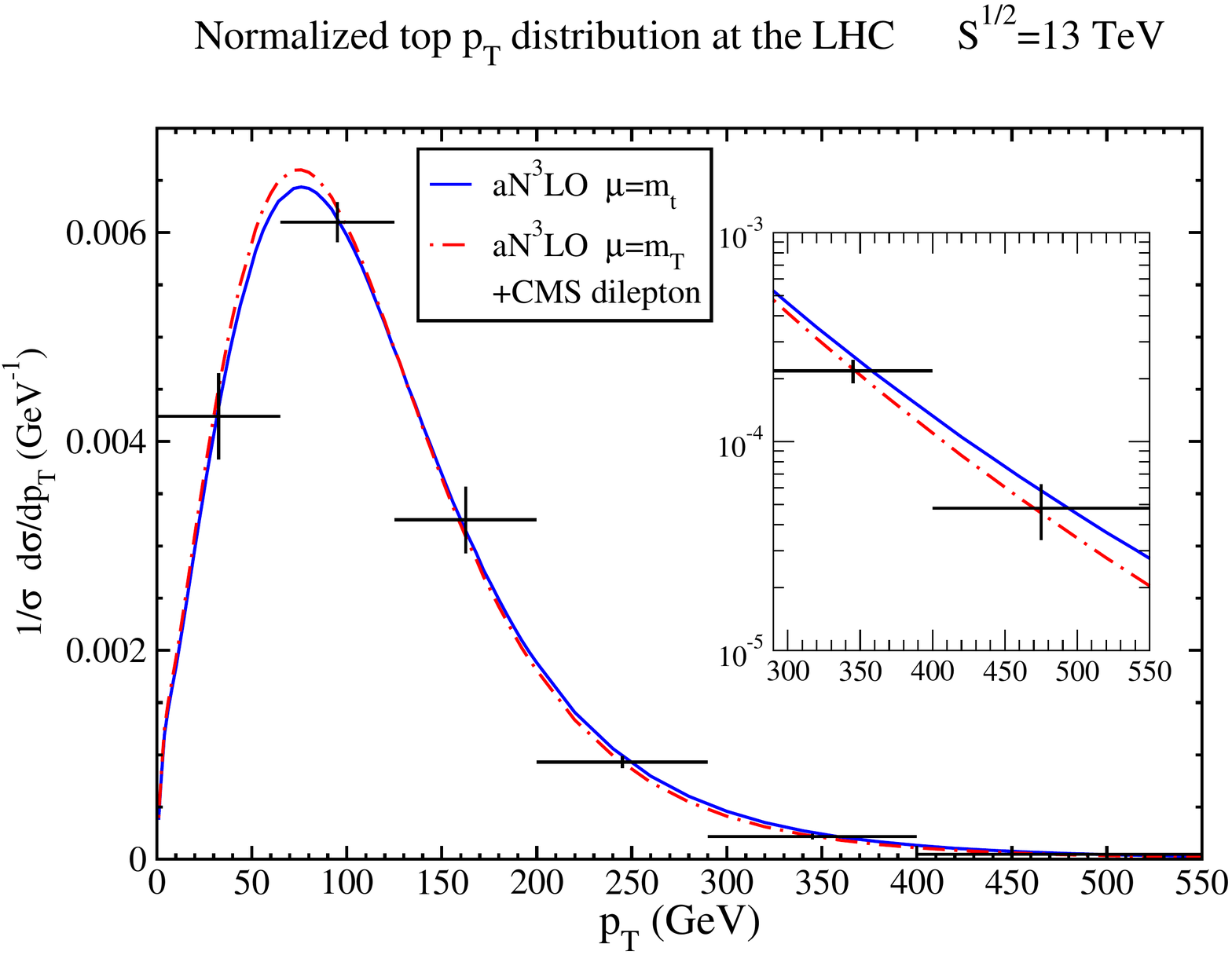}
\hspace{-3mm}
\includegraphics[width=85mm]{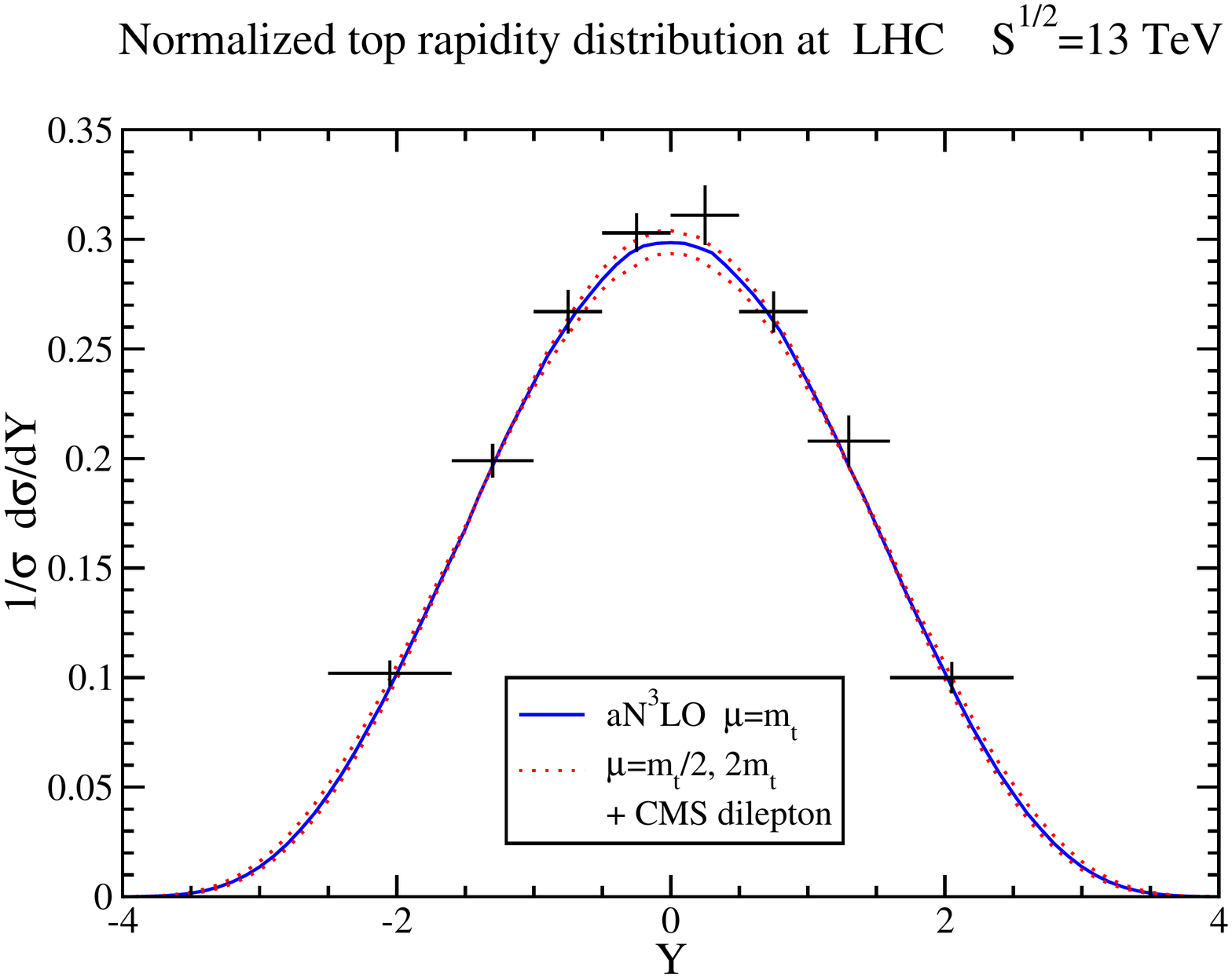}
\caption{aN$^3$LO top-quark normalized $p_T$ (left) and rapidity (right) distributions at 13 TeV energy compared with CMS \cite{CMSpty13} data.}
\label{ttpTY}
\end{figure}

Top-quark differential distributions can provide a lot more information than total cross sections, and they are sensitive to new physics. Top-quark transverse-momentum ($p_T$) and rapidity distributions have been calculated to aN$^3$LO, and the soft-gluon contributions are very important \cite{NKtop,QCDwork18,NKtt}. 

The aN$^3$LO top-quark normalized $p_T$ distributions, $(1/\sigma) d\sigma/dp_T$, and normalized rapidity distributions, $(1/\sigma) d\sigma/dY$,  are shown in Fig. \ref{ttpTY} at 13 TeV energy and compared with CMS data. Again, we find excellent agreement of the aN$^3$LO theoretical predictions for both distributions with the corresponding data.

\mysection{Summary}

We have presented results with soft-gluon corrections for total cross sections and differential distributions for various processes involving the production of charged Higgs bosons and top quarks. The soft-gluon corrections are significant and dominant in all the processes that we have discussed. 

We have presented aN$^3$LO results for $tH^-$ production, for $tW$ production, and for $t{\bar t}$ production. We have shown aNNLO results for $t$-channel and $s$-channel single-top production, $H^-W^+$ production, and $tZ$ production via anomalous couplings. For the single-top and top-antitop-pair processes, we find excellent agreement with all available collider data. The higher-order soft-gluon corrections are needed for a better description of the data and for setting limits in new physics searches.

\mysection*{Acknowledgments}

This material is based upon work supported by the National Science Foundation under Grant No. PHY 1519606.


\begin{thebibliography}{}

\bibitem{NKtop}
N. Kidonakis, Int. J. Mod. Phys. A \textbf{33}, 1830021 (2018) [arXiv:1806.03336 [hep-ph]].

\bibitem{NKtH}
N. Kidonakis, JHEP \textbf{0505}, 011 (2005) [hep-ph/0412422]; 
Phys. Rev. D \textbf{94}, 014010 (2016) [arXiv:1605.00622 [hep-ph]]; 
in Proceedings of CIPANP2018 [arXiv:1808.02935 [hep-ph]].

\bibitem{MMHT2014}
L.A. Harland-Lang, A.D. Martin, P. Molytinski, and R.S. Thorne,   
Eur. Phys. J. C \textbf{75}, 204 (2015) [arXiv:1412.3989 [hep-ph]].

\bibitem{NKHW}
N. Kidonakis, Phys. Rev. D \textbf{97}, 034002 (2018) [arXiv:1704.08549 [hep-ph]].

\bibitem{NKtZ}
N. Kidonakis, Phys. Rev. D \textbf{97}, 034028 (2018) [arXiv:1712.01144 [hep-ph]].

\bibitem{NLOtZ}
B.H. Li, Y. Zhang, C.S. Li, J. Gao, and H.X. Zhu, Phys. Rev. D \textbf{83}, 114049 (2011) [arXiv:1103.5122 [hep-ph]]. 

\bibitem{CT14}
S. Dulat, T.-J. Hou, J. Gao, M. Guzzi, J. Huston, P. Nadolsky, J. Pumplin, C. Schmidt, D. Stump, and C.-P. Yuan, Phys. Rev. D \textbf{93}, 033006 (2016) [arXiv:1506.07443 [hep-ph]].

\bibitem{CMStqZ}
CMS Collaboration, JHEP \textbf{1707}, 003 (2017) [arXiv:1702.01404 [hep-ex]].

\bibitem{ATLAStqZ}
ATLAS Collaboration, ATLAS-CONF-2017-070.

\bibitem{MFNK}
M. Forslund and N. Kidonakis, arXiv:1808.09014 [hep-ph].

\bibitem{NNLOtch}
M. Brucherseifer, F. Caola, and K. Melnikov,  
Phys. Lett. B \textbf{736}, 58 (2014) [arXiv:1404.7116 [hep-ph]].

\bibitem{BGYZ}
E.L. Berger, J. Gao, C.-P. Yuan, and H.X. Zhu, 
Phys. Rev. D \textbf{94}, 071501 (2016) [arXiv:1606.08463 [hep-ph]].

\bibitem{BGZ}
E.L. Berger, J. Gao, and H.X. Zhu, 
JHEP \textbf{1711}, 158 (2017) [arXiv:1708.09405 [hep-ph]].

\bibitem{NNLOsch}
Z. Liu and J. Gao, arXiv:1807.03835 [hep-ph].

\bibitem{Zhu}
S.H. Zhu, Phys. Lett. B \textbf{524}, 283 (2002) [Erratum: \textit{ibid.} 
\textbf{537}, 351 (2002)] [hep-ph/0109269].

\bibitem{NKsingletop}
N. Kidonakis, Phys. Rev. D \textbf{81}, 054028 (2010) [arXiv:1001.5034 [hep-ph]];
\textbf{82}, 054018 (2010) [arXiv:1005.4451 [hep-ph]];
\textbf{83}, 091503(R) (2011) [arXiv:1103.2792 [hep-ph]].

\bibitem{NKtW16} 	
N. Kidonakis, Phys. Rev. D \textbf{96}, 034014 (2017) [arXiv:1612.06426 [hep-ph]].

\bibitem{CMStch7}
CMS Collab., JHEP \textbf{1212}, 035 (2012) [arXiv:1209.4533 [hep-ex]].

\bibitem{ATLAStch7}
ATLAS Collab., Phys. Rev. D \textbf{90}, 112006 (2014) [arXiv:1406.7844 [hep-ex]].

\bibitem{CMStch8}
CMS Collab., JHEP \textbf{1406}, 090 (2014) [arXiv:1403.7366 [hep-ex]]. 

\bibitem{ATLAStch8}
ATLAS Collab., Eur. Phys. J. C \textbf{77}, 531 (2017) [arXiv:1702.02859 [hep-ex]].  

\bibitem{ATLAStch13}
ATLAS Collab., JHEP \textbf{1704}, 086 (2017) [arXiv:1609.03920 [hep-ex]]. 

\bibitem{CMStch13}
CMS Collab., Phys. Lett. B \textbf{772}, 752 (2017) [arXiv:1610.00678 [hep-ex]].

\bibitem{CDFD0tch}
CDF and D0 Collab., Phys. Rev. Lett. \textbf{115}, 152003 (2015) [arXiv:1503.05027 [hep-ex]].

\bibitem{ATLAStW7} 
ATLAS Collab., Phys. Lett. B \textbf{716}, 142 (2012) [arXiv:1205.5764 [hep-ex]].

\bibitem{CMStW7} 
CMS Collab., Phys. Rev. Lett. \textbf{110}, 022003 (2013) [arXiv:1209.3489 [hep-ex]].

\bibitem{ATLASCMStW8}
ATLAS and CMS Collab., ATLAS-CONF-2016-023, CMS-PAS-TOP-15-019.

\bibitem{ATLAStW13}
ATLAS Collab., JHEP \textbf{1801}, 063 (2018) [arXiv:1612.07231 [hep-ex]].

\bibitem{CMStW13}
CMS Collab., arXiv:1805.07399 [hep-ex].

\bibitem{QCDwork18}
N. Kidonakis, in Proceedings of QCD@Work 2018 [arXiv:1809.02524 [hep-ph]].

\bibitem{NKtt}
N. Kidonakis, Phys. Rev. D \textbf{90}, 014006 (2014) [arXiv:1405.7046 [hep-ph]]; \textbf{91}, 031501(R) (2015) [arXiv:1411.2633 [hep-ph]]; 
\textbf{91}, 071502(R) (2015) [arXiv:1501.01581 [hep-ph]].

\bibitem{CMS5.02}
CMS Collab., JHEP \textbf{1803}, 115 (2018) [arXiv:1711.03143 [hep-ex]]. 

\bibitem{ATLAStt7}
ATLAS Collab.,  Eur. Phys. J. C \textbf{74}, 3109 (2014) [Addendum: \textit{ibid.} {\bf 76}, 642 (2016)] [arXiv:1406.5375 [hep-ex]].

\bibitem{CMStt7and8}
CMS Collab., JHEP \textbf{1608}, 029 (2016) [arXiv:1603.02303 [hep-ex]]. 

\bibitem{ATLAStt8}
ATLAS Collab., Eur. Phys. J. C \textbf{78}, 487 (2018) [arXiv:1712.06857 [hep-ex]].	

\bibitem{ATLAStt13}
ATLAS Collab., Phys. Lett. B \textbf{761}, 136 (2016) [Erratum: \textit{ibid.}
\textbf{772}, 879 (2017)] [arXiv:1606.02699 [hep-ex]]. 

\bibitem{CMStt13}
CMS Collab., JHEP \textbf{1709}, 051 (2017) [arXiv:1701.06228 [hep-ex]]. 

\bibitem{CDF1.8}
CDF Collab., Phys. Rev. D \textbf{64}, 032002 (2001) [Erratum: \textit{ibid.} \textbf{67}, 119901 (2003)] [hep-ex/0101036].

\bibitem{D01.8}
D0 Collab., Phys. Rev. D \textbf{67}, 012004 (2003) [hep-ex/0205019].

\bibitem{CDFD01.96}
CDF and D0 Collab., Phys. Rev. D \textbf{89}, 072001 (2014) [arXiv:1309.7570 [hep-ex]].

\bibitem{CMSpty13}
CMS Collab., JHEP \textbf{1804}, 060 (2018) [arXiv:1708.07638 [hep-ex]]. 


\end{thebibliography}
\end{document}